\documentstyle[twoside,fleqn,espcrc2,psfig,epsfig]{article}

\def\eq#1{{eq.~(\ref{#1})}}

\def\beq{\begin{equation}}
\def\eeq{\end{equation}}
\def\lsim{\raise0.3ex\hbox{$\;<$\kern-0.75em\raise-1.1ex\hbox{$\sim\;$}}}
\def\gsim{\raise0.3ex\hbox{$\;>$\kern-0.75em\raise-1.1ex\hbox{$\sim\;$}}}

\def\apj#1#2#3{ Astrophys. J. #1 (19#2) #3.}

\def\np#1#2#3{ Nucl. Phys. #1 (19#2) #3.}

\def\pl#1#2#3{ Phys. Lett. #1 (19#2) #3.}
\def\pr#1#2#3{ Phys. Rev. #1 (19#2) #3.}

\def\prl#1#2#3{ Phys. Rev. Lett. #1 (19#2) #3.}

\def\rpp#1#2#3{ Rept. Prog. Phys. #1 (19#2) #3.}

\def\Frac#1#2{\frac{\displaystyle{#1}}{\displaystyle{#2}}}

\newcommand{\AmS}{{\protect\the\textfont2
  A\kern-.1667em\lower.5ex\hbox{M}\kern-.125emS}}

\title{Relic neutrino asymmetry, CMB and large-scale structure}
\author{Sergio Pastor\thanks{Speaker, e-mail: pastor@sissa.it}$~^{\rm a}$
and 
Julien Lesgourgues\thanks{E-mail: lesgour@sissa.it.
Work supported by INFN and by the TMR network grant
ERBFMRXCT960090.}
\address{SISSA--ISAS and INFN, Sezione di Trieste\\
Via Beirut 2-4, I-34013 Trieste, Italy}}

\begin{document}

\begin{abstract}
We consider some consequences of the presence of a cosmological lepton
asymmetry in the form of neutrinos. A relic neutrino degeneracy enhances
the contribution of massive neutrinos to the present energy density of
the Universe, and modifies the power spectrum of radiation and matter.
Comparing with current observations of cosmic microwave background
anisotropies and large scale structure, we derive some constraints on
the relic neutrino degeneracy and on the spectral index in the case of a
flat Universe with a cosmological constant.
\end{abstract}


\maketitle

\section{Introduction}

It is generally assumed that our Universe contains an approximately
equal amount of leptons and antileptons. The lepton asymmetry would be
of the same order as the baryon asymmetry, which is very small as
required by Big Bang Nucleosynthesis (BBN) considerations.  The
existence of a large lepton asymmetry is restricted to be in the form
of neutrinos from the requirement of universal electric neutrality,
and the possibility of a large neutrino asymmetry is still open. {}From
a particle physics point of view, a lepton asymmetry can be generated
by an Affleck-Dine mechanism \cite{AF} without producing a large
baryon asymmetry (see ref.~\cite{Casas} for a recent model), or even
by active-sterile neutrino oscillations after the electroweak phase
transition \cite{Foot}. 

We have studied some cosmological implications of relic degenerate
neutrinos \cite{Paper} (here degenerate refers to
neutrino-antineutrino asymmetry, not to mass degeneracy). We do not
consider any specific model for generating such an asymmetry, and just
assume that it was created well before neutrinos decouple from the
rest of the plasma.  An asymmetry of order one or larger can have
crucial effects on the global evolution of the Universe. Among other
effects, it changes the decoupling temperature of neutrinos, the
primordial production of light elements at BBN, the time of equality
between radiation and matter, or the contribution of relic neutrinos
to the present energy density of the Universe. The latter changes
affect the evolution of perturbations in the Universe. We focus on the
anisotropies of the Cosmic Microwave Background (CMB), and on the
distribution of Large Scale Structure (LSS). We calculate the power
spectrum of both quantities, in the case of massless degenerate
neutrinos, and also for neutrinos with a mass of $0.07$ eV, as
suggested to explain the experimental evidence of atmospheric neutrino
oscillations at Super-Kamiokande \cite{SK}.

The effect of neutrino degeneracy on the LSS power spectrum was
studied in ref.~\cite{Larsen}, as a way of improving the agreement
with observations of mixed dark matter models with eV neutrinos, in
the case of high values of the Hubble parameter.  Adams \& Sarkar
\cite{Sarkar} calculated the CMB anisotropies and the matter power
spectrum, and compared them with observations in the
$\Omega_\Lambda=0$ case for massless degenerate neutrinos. More
recently, Kinney \& Riotto \cite{Kinney} also calculated the CMB
anisotropies for massless degenerate neutrinos in the
$\Omega_\Lambda=0.7$ case.

\vspace{-0.25cm}
\section{Energy density of massive degenerate neutrinos}
\label{energy}

The energy density of one species of massive degenerate neutrinos and
antineutrinos, described by the distribution functions $f_\nu$ and
$f_{\bar{\nu}}$, is (we use $\hbar=c=k_B=1$ units)
\beq
\rho_\nu \! + \! \rho_{\bar{\nu}}= \!\!
\int_0^\infty \!\!\! \frac{dp}{2\pi^2} ~p^2 \sqrt{p^2 \! + \! m_\nu^2}
(f_\nu(p) \! + \! f_{\bar{\nu}}(p))
\label{defrhonu}
\eeq
valid at any moment. Here $p$ is the magnitude of the 3-momentum and
$m_\nu$ is the neutrino mass.

When the early Universe was hot enough, the neutrinos were in
equilibrium with the rest of the plasma via the weak interactions. In
that case the distribution functions $f_\nu$ and $f_{\bar{\nu}}$
changed with the Universe expansion, keeping the form of a Fermi-Dirac
distribution,
\beq 
f_{\nu,\bar{\nu}}(p)=\Frac{1}{\exp \left(\frac{p}{T_\nu} \mp
\frac{\mu}{T_\nu}\right)+1}
\label{FD}
\eeq
Here $\mu$ is the neutrino chemical potential, which is nonzero if a
neutrino-antineutrino asymmetry has been previously produced. Later
the neutrinos decoupled when they were still relativistic, and from
that moment the neutrino momenta just changed according to the
cosmological redshift.  If $a$ is the expansion factor, the neutrino
momentum decreases keeping $ap$ constant.  At the same time the
neutrino degeneracy parameter $\xi \equiv \mu/T_\nu$ is conserved,
with a value equal to that at the moment of decoupling. Therefore one
can still calculate the energy density of neutrinos now from
\eq{defrhonu} and \eq{FD}, replacing $\mu/T_\nu$ by $\xi$ and
$p/T_\nu$ by $p/(y_\nu T_0)$, where $T_0 \simeq 2.726$ K and $y_\nu$
is the present ratio of neutrino and photon temperatures, which is not
unity because once decoupled the neutrinos did not share the entropy
transfer to photons from the successive particle annihilations that
occurred in the early Universe. 

In the presence of a significant neutrino degeneracy $\xi$ the
decoupling temperature $T(\xi)$ is higher than in the standard case,
\cite{Freese,Kang}. The reaction rate $\Gamma$ of the weak processes,
that keep the neutrinos in equilibrium with the other species, is
reduced because some of the initial or final neutrino states will be
occupied. The authors of ref.~\cite{Kang} found that the neutrino
decoupling temperature is $T_{dec}(\xi) \approx
0.2\xi^{2/3}\exp(\xi/3)$ MeV (for $\nu_\mu$ or $\nu_\tau$). Therefore
if $\xi$ is large enough, the degenerate neutrinos decouple before the
temperature of the Universe drops below the different mass thresholds,
and are not heated by the particle-antiparticle annihilations,
reducing the ratio of neutrino and photon temperatures with respect
to the standard value $y_\nu=(4/11)^{1/3}$.

\begin{figure}[htb]
\vspace{-0.5cm}
\centerline{\psfig{file=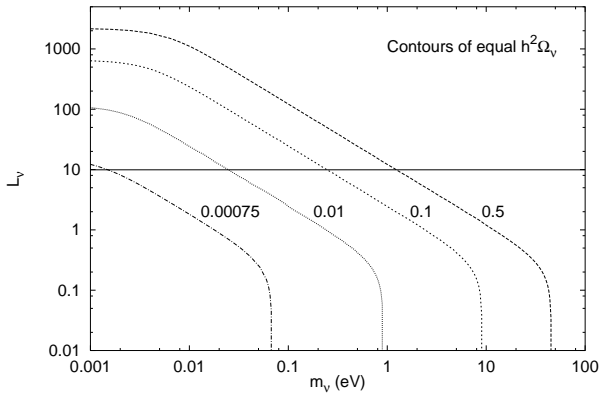,angle=-90,width=0.49\textwidth}}
\vspace{-0.75cm}
\caption{Present energy density of massive degenerate neutrinos as a
function of the neutrino asymmetry.}
\label{lnumass}
\end{figure}

The present contribution of these degenerate neutrinos to the energy
density of the Universe can be parametrized as $\rho_\nu = 10^4 h^2
\Omega_\nu$ eV cm$^{-3}$, where $\Omega_\nu$ is the neutrino energy
density in units of the critical density $\rho_c=3H^2M_P^2/8\pi$,
$M_P=1.22 \times 10^{19}$ GeV is the Planck mass and $H=100h$ Km
s$^{-1}$ Mpc$^{-1}$ is the Hubble parameter.
The value of $\rho_\nu$ can be calculated as a function of the
neutrino mass and the neutrino degeneracy $\xi$, or equivalently the
present neutrino asymmetry $L_\nu$ defined as the following ratio of
number densities
\beq 
L_\nu \equiv \frac{n_\nu-n_{\bar{\nu}}}{n_\gamma} =
\frac{1}{12\zeta (3)} y^3_\nu [\xi^3 + \pi^2 \xi]
\label{Lnu}
\eeq
We show\footnote{Here we assume $\xi>0$, but the results are also
valid for $\xi<0$ provided that $\xi$ and $L_\nu$ are understood as
moduli.} in figure \ref{lnumass} the contours in the $(m_\nu,L_\nu)$
plane that correspond to some particular values of $h^2
\Omega_\nu$. In the limit of small degeneracy (vertical lines) one
recovers the well-known bound on the neutrino mass $m_\nu \lsim 46$ eV
for $h^2 \Omega_\nu=0.5$. On the other hand, for very light
neutrinos the horizontal lines set a maximum value on the neutrino
degeneracy, that would correspond to a present neutrino chemical
potential $\mu_0 \lsim 7.4 \times 10^{-3}$ eV, also for $h^2
\Omega_\nu=0.5$. In the intermediate region of the figures the
neutrino energy density is $\rho_\nu \simeq m_\nu n_\nu (\xi)$ and the
contours follow roughly the relation
$L_\nu (m_\nu/\mbox{eV})\simeq 24.2 h^2\Omega_\nu$.

A similar calculation has been recently performed in reference
\cite{PalKar}. Note however that the ratio of neutrino and
photon temperatures was not properly taken into account for large
$\xi$.

The presence of a neutrino degeneracy can modify the outcome of BBN
(for a review see \cite{Sarkar96}). First a larger neutrino energy
density increases the expansion rate of the Universe, thus enhancing
the primordial abundance of $^4$He. This is valid for a nonzero $\xi$
of any neutrino flavor.  In addition if the degenerate neutrinos are
of electron type, they have a direct influence over the weak processes
that interconvert neutrons and protons. This last effect depends on
the sign of $\xi_{\nu_e}$, and one gets $-0.06 \lsim
\xi_{\nu_e} \lsim 1.1$ \cite{Kang},
%
%
while a sufficiently long matter dominated epoch requires
$|\xi_{\nu_\mu,\nu_\tau}| \lsim 6.9$ \cite{Kang}.  This estimate
agrees with our analysis in section \ref{comparison} and places a
limit shown by the horizontal line in figure \ref{lnumass} in
the case of degenerate $\nu_\mu$ or $\nu_\tau$.

\section{Effects on the power spectra}
\label{results}

\begin{figure*}[t]
\vspace{-0.5cm}
\begin{eqnarray}
\psfig{file=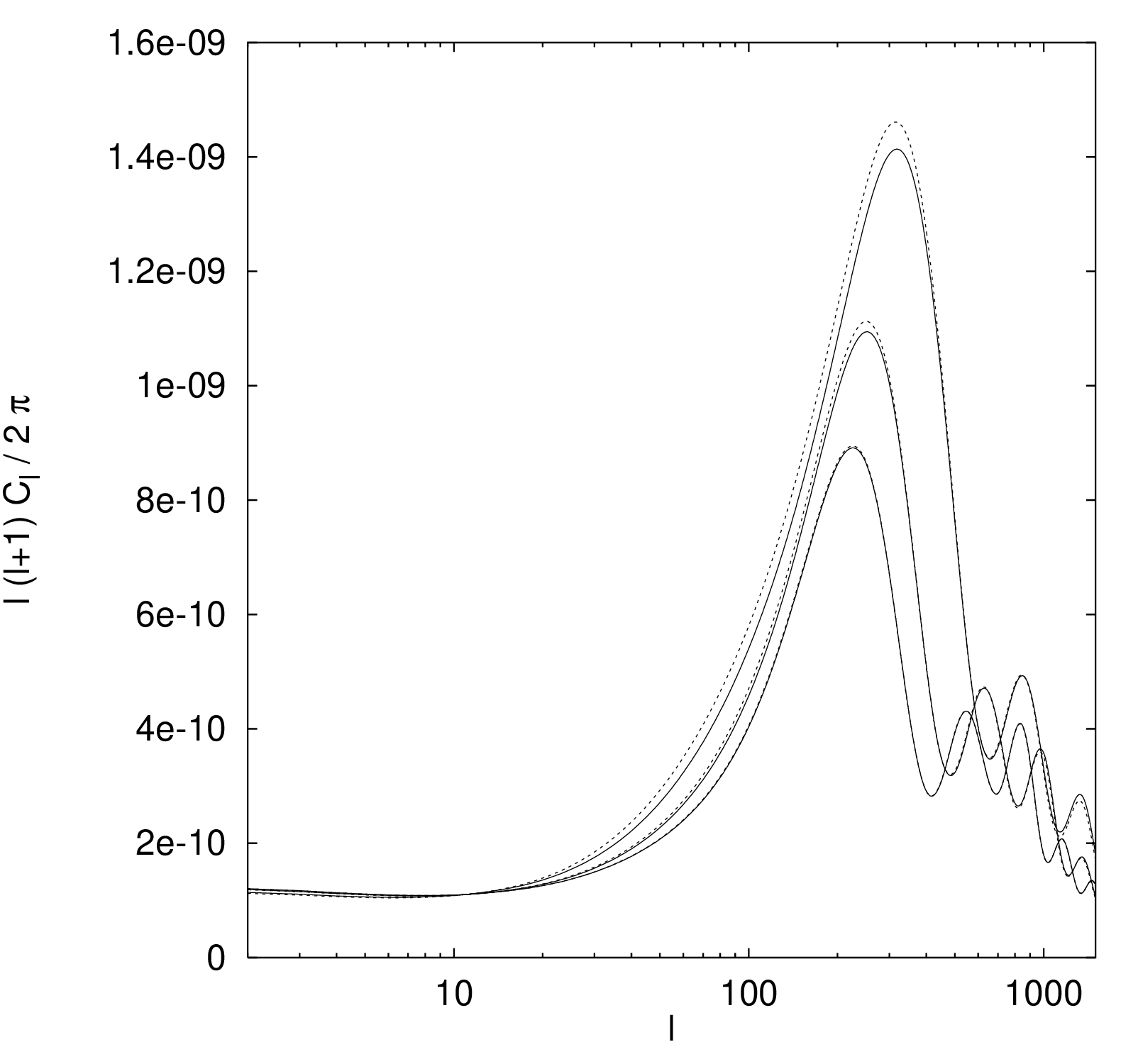,width=0.48\textwidth}~~~~~
\psfig{file=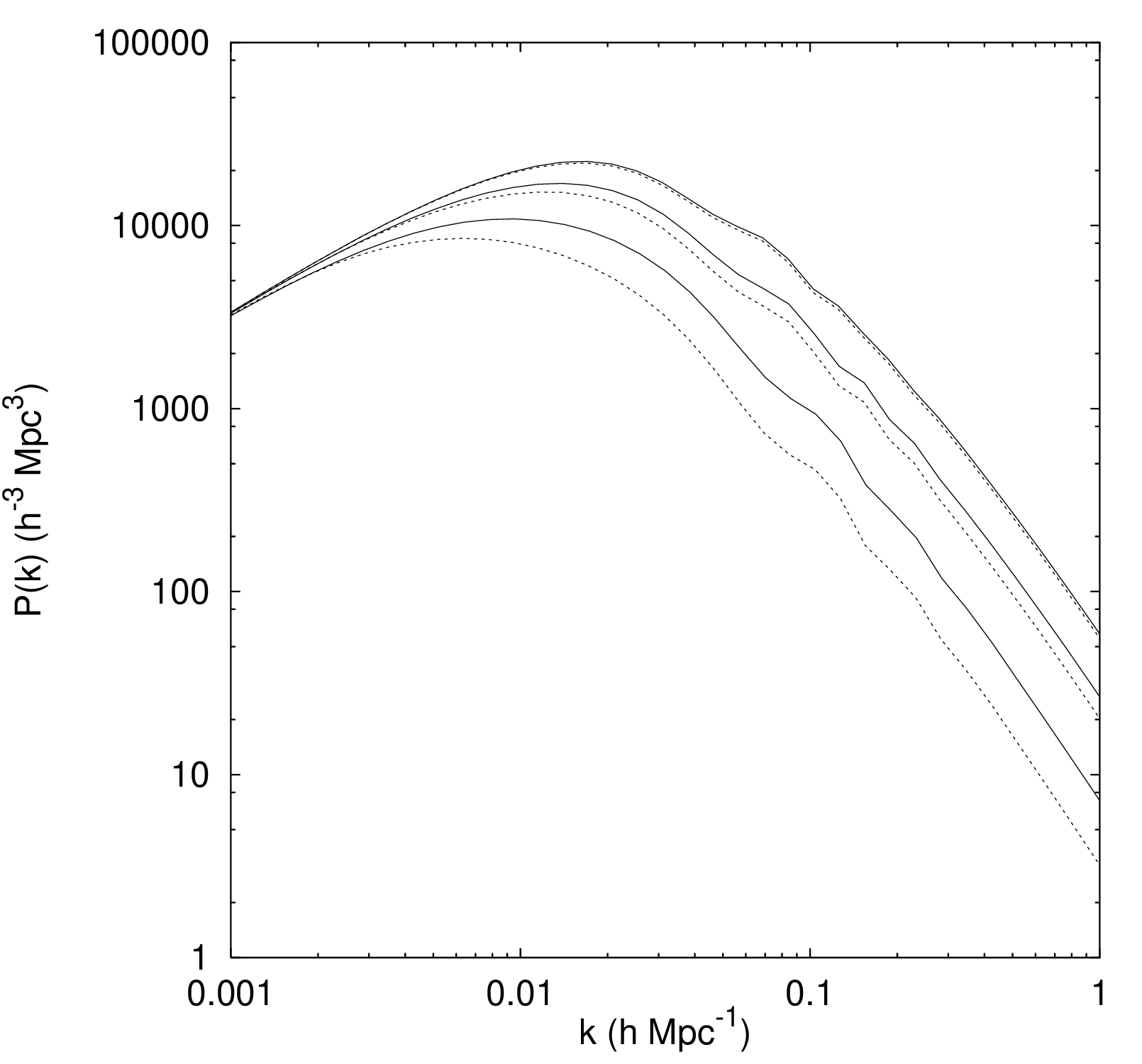,width=0.48\textwidth}
\nonumber
\end{eqnarray}
\vspace{-1.5cm}
\caption{CMB anisotropy and matter power spectra
for different models with one family of massless (solid lines) and
$m_{\nu} = 0.07$ eV (dashed lines) degenerate neutrinos. From bottom
to top (from top to bottom for $P(k)$), $\xi=0,3,5$.  
Cosmological parameters are fixed as described in the text. 
}
\label{fig.CMB}
\end{figure*}
%
%

We compute the power spectra of CMB anisotropies and LSS
using the Boltzmann code {\tt cmbfast} by Seljak \&
Zaldarriaga \cite{SelZal}, adapted to the case of one family of
degenerate neutrinos ($\nu$, $\bar{\nu}$), with mass $m_\nu$ and
degeneracy parameter $\xi$. Our modifications to the code
are reviewed and explained in \cite{Paper}.  

The effect of $\xi$ and $m_{\nu}$ on the CMB anisotropy spectrum can
be seen in figure \ref{fig.CMB}. We choose a set of cosmological
parameters ($h=0.65$, $\Omega_b=0.05$, $\Omega_{\Lambda}=0.70$,
$\Omega_{CDM}=1-\Omega_b-\Omega_{\nu}-\Omega_{\Lambda}$,
$Q_{rms-ps}=18~\mu$K, flat primordial spectrum, no reionization, no
tensor contribution), and we vary $\xi$ from 0 to 5, both in the case
of massless degenerate neutrinos and degenerate
neutrinos with $m_{\nu}=0.07$ eV.

Let us first comment the massless case.  The main effect of $\xi$ is
to boost the amplitude of the first peak\footnote{In fact, this is not
true for very large values of $\xi$, where recombination can take
place still at the end of radiation domination, and anisotropies are
suppressed.  However in such a case the location of the first peak is
$l \gsim 450$, and the matter power spectrum is strongly
suppressed.}. Indeed, increasing the energy density of radiation
delays matter-radiation equality, which is known to boost the acoustic
peaks, and to shift them to higher multipoles, by a factor $( (1 +
a_{eq}/a_*)^{1/2} - (a_{eq}/a_*)^{1/2})^{-1}$ ($a_{eq}$ increases with
$\xi$, while the recombination scale factor $a_*$ is almost
independent of the radiation energy density). Secondary peaks are then
more affected by diffusion damping at large $l$, and their amplitude
can decrease with $\xi$.

In the case of degenerate neutrinos with $m_{\nu}=0.07$ eV, the
results are quite similar in first approximation. Indeed, the effects
described previously depend on the energy density of neutrinos at
equality. At that time, they are still relativistic, and identical to
massless neutrinos with equal degeneracy parameter.  However, with a
large degeneracy, $\Omega_{\nu}$ today becomes significant: for
$\xi=5$, one has $\Omega_{\nu}=0.028$, i.e. the same order of
magnitude as $\Omega_b$.  Since we are studying flat models,
$\Omega_{\nu}$ must be compensated by less baryons, cold dark matter
(CDM) or $\Omega_{\Lambda}$. In our example, $\Omega_b$ and
$\Omega_{\Lambda}$ are fixed, while $\Omega_{CDM}$ slightly
decreases. This explains the small enhancement of the first peak
compared to the massless case (3.4\% for $\xi=5$).  Even if this
effect is indirect, it is nevertheless detectable in principle,
possibly by future satellite missions {\it MAP} and {\it Planck} (even
if one does not impose the flatness condition, the effect of
$\Omega_{\nu}$ will be visible through a modification of the
curvature).


We also plot in figure \ref{fig.CMB} the power spectrum $P(k)$,
normalized on large scales to COBE. The effect of both parameters
$\xi$ and $m_{\nu}$ is now to suppress the power on small scales.
Indeed, increasing $\xi$ postpones matter-radiation equality, allowing
less growth for fluctuations crossing the Hubble radius during
radiation domination. Adding a small mass affects the recent evolution
of fluctuations, and has now a direct effect: when the degenerate
neutrinos become non-relativistic, their free-streaming suppresses
the growth of fluctuations for scales within the Hubble radius.  This
effect, already known for non-degenerate neutrinos \cite{Huetal}, is
enhanced in the presence of a neutrino degeneracy, since the average
neutrino momentum is shifted to larger values.


Our results for massless degenerate neutrinos can be compared with
those of previous works. We found the same effect of $\xi$ on the CMB
for $\Omega_{\Lambda}=0$ as in \cite{Sarkar}, while the revised
results in \cite{Kinney} also agree 
with our calculations for $\Omega_{\Lambda}=0.7$.

\section{Comparison with observations}

\label{comparison}

Since the degeneracy increases dramatically the amplitude of the first
CMB peak, we expect large $\xi$ values to be favored in the case of
cosmological models known to predict systematically a low peak (unless
a large blue tilt is invoked, which puts severe constraints on
inflation).
Our goal here is not to explore systematically all possibilities, but
to briefly illustrate how $\xi$ can be constrained by current
observations for flat models with different values of
$\Omega_{\Lambda}$. Recent results from supernovae, combined with CMB
constraints, favor flat models with $\Omega_{\Lambda} \sim 0.6-0.7$.

We choose a model with $h=0.65$, $\Omega_b=0.05$,
$Q_{rms-ps}=18~\mu$K, no reionization and no tensor contribution, and
look for the allowed window in the space of free parameters
($\Omega_{\Lambda},\xi,n$).  The allowed window is defined as the
intersection of regions preferred at the 95\% confidence level by four
independent experimental tests, based on $\sigma_8$ estimation,
Stromlo-APM redshift survey, bulk velocity reconstruction, and CMB
anisotropy measurements. Details concerning these tests can be found
in \cite{Paper}.

%

\begin{figure*}[t]
\vspace{-0.5cm}
\begin{eqnarray}
\psfig{file=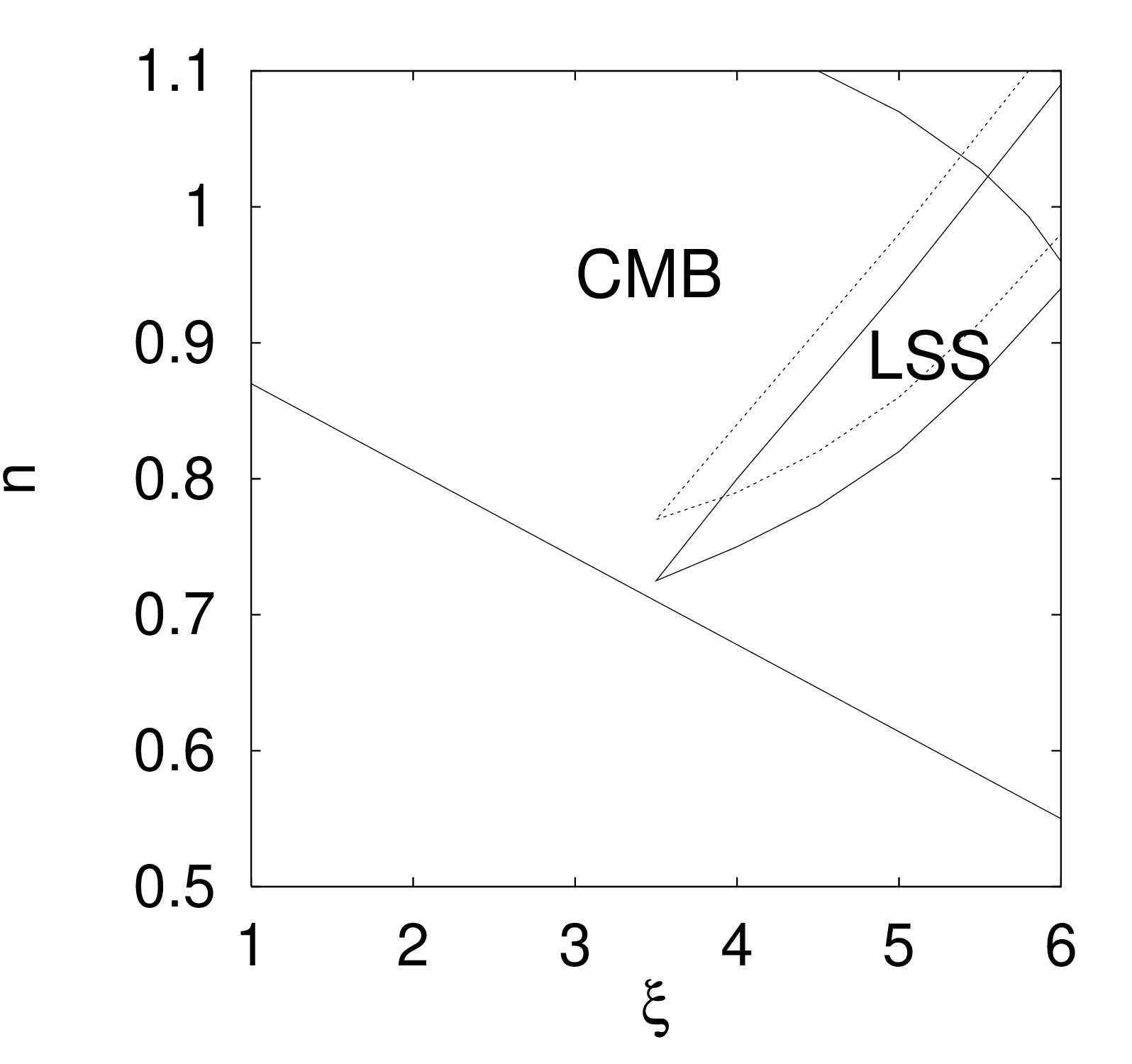,width=0.47\textwidth}~~~~
\psfig{file=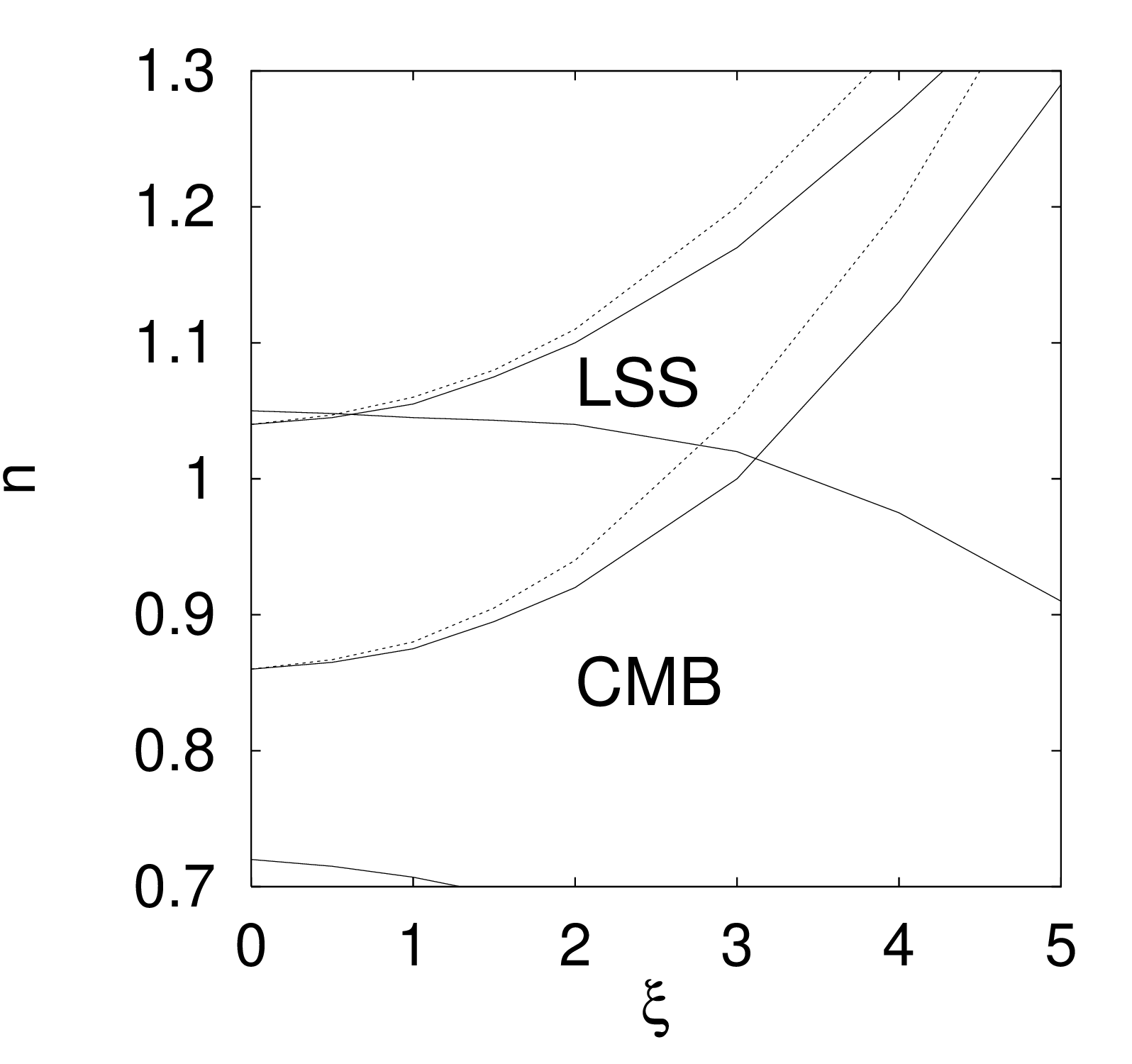,width=0.47\textwidth}
\nonumber
\end{eqnarray}
\vspace{-1.5cm}
\caption{LSS and CMB constraints in ($\xi$, $n$) space
for $\Omega_{\Lambda}=0$ (left) and $\Omega_{\Lambda}=0.6$ (right).
The underlying cosmological model is flat, with $h=0.65$,
$\Omega_b=0.05$, $Q_{rms-ps}=18~\mu$K, no reionization, no tensor
contribution. The allowed regions are those where the labels are. For
LSS constraints, we can distinguish between degenerate neutrinos with
$m_{\nu} =0$ (solid lines) and $m_{\nu} =0.07$ eV (dotted lines).}
\label{fig.WIN}
\end{figure*}
%

We plot in figure \ref{fig.WIN} the LSS and CMB allowed regions in
($\xi$, $n$) parameter space, for $\Omega_{\Lambda}=0$
and $0.6$.  
In the case of degenerate neutrinos with $m_{\nu} = 0.07$ eV, the LSS
regions are slightly shifted at large $\xi$, since, as we saw, the
effect of $\xi$ is enhanced (dotted lines on the figure). The CMB
regions do not show this distinction, given the smallness of the
effect and the imprecision of the data.  One can immediately see that
LSS and CMB constraints on $n$ are shifted in opposite direction with
$\xi$: indeed, the effects of $\xi$ and $n$ both produce a higher CMB
peak, while to a certain extent they compensate each other in $P(k)$.
So, for $\Omega_{\Lambda}\geq0.7$, a case in which a power spectrum
normalized to both COBE and $\sigma_8$ yields a too high
peak\footnote{At least, for the values of the other cosmological
parameters considered here.  This situation can be easily improved,
for instance, with $h=0.7$.}, a neutrino degeneracy can only make
things worst, and we find no allowed window at all.  In the other
extreme case $\Omega_{\Lambda}=0$, it is well known that the amplitude
required by $\sigma_8$ and the shape probed by redshift surveys favor
different values of $n$. We find that the neutrino degeneracy can
solve this problem with $\xi \gsim 3.5$, but the allowed window
is cut at $\xi \simeq 6$ by CMB data, and we are left with an
interesting region in which $\Omega_0=1$ models are viable. This
result is consistent with \cite{Sarkar}. However, current evidences
for a low $\Omega_0$ Universe are independent of the constraints used
here, so there are not many motivations at the moment to consider this
window seriously.  Finally, for $\Omega_{\Lambda}=0.5-0.6$, a good
agreement is found up to $\xi \simeq 3$. This upper bound could
marginally explain the generation of ultra-high energy cosmic rays by
the annihilation of high-energetic neutrinos on relic neutrinos with
mass $m_{\nu}= 0.07$ eV \cite{Gelmini}.

\end{document}